%% file: main.tex
\documentclass[sigconf]{acmart}

\usepackage{booktabs} 
\usepackage{epstopdf}
\usepackage{tabularx}
\usepackage{multirow}
\pdfmapfile{+txfonts.map}
\usepackage{multirow}
\usepackage{amsmath}
\usepackage{algorithm2e}
\usepackage{caption}
\usepackage{subcaption}

\setcopyright{rightsretained}

\usepackage{amssymb}
\usepackage{pifont}
\newcommand{\cmark}{\ding{51}}%
\newcommand{\xmark}{\ding{55}}%

\acmDOI{10.475/123_4}

\acmISBN{123-4567-24-567/08/06}

\acmYear{2019}
\copyrightyear{2019}

\acmPrice{15.00}

\begin{document}
\title{Latent Unexpected and Useful Recommendation}
\author{Pan Li}
\affiliation{\institution{New York University, \\Stern School of Business}}
\email{pli2@stern.nyu.edu}

\author{Alexander Tuzhilin}
\affiliation{\institution{New York University, \\Stern School of Business}}
\email{atuzhili@stern.nyu.edu}

\renewcommand{\shortauthors}{}
\renewcommand{\shorttitle}{}
\begin{abstract}
Providing unexpected recommendations is an important task for recommender systems. To do this, we need to start from expectations of users and deviate from these expectations when recommending items. Previously proposed approaches model user expectations in the feature space, making them limited to the items that the user has visited or expected by the deduction of associate rules, without including the items that the user could also expect from the latent, complex and heterogeneous interactions between users, items and entities. In this paper, we define unexpectedness in the latent space rather than in the feature space and develop a novel Latent Convex Hull (LCH) method to provide unexpected recommendations. Extensive experiments on two real-world datasets demonstrate effectiveness of the proposed model that significantly outperforms alternative state-of-the-art unexpected recommendation methods in terms of unexpectedness measures while achieving the same level of accuracy.

\end{abstract}



\maketitle

\input{Introduction}
\input{RelatedWork}
\input{Method}
\input{Experiments}
\input{Conclusions}

\bibliographystyle{ACM-Reference-Format}
\bibliography{sigproc}
\end{document}

%% file: Introduction.tex
\section{Introduction}
Recommender systems have been playing an important role in the process of information dissemination and online commerce, which assist the user in filtering for the best content while shaping their consumption behavior patterns at the same time.  However, many classical recommender systems are facing the problem of a filter bubble \cite{pariser2011filter,nguyen2014exploring}, which means that the targeted users would only get recommendations of a small portion of the available items, and they tend to get more recommendations of the items that they are most familiar with. For example, a Harry Potter fan may feel unsatisfied if the system keeps recommending Harry Potter series. This type of filter bubble phenomenon has motivated researchers to introduce several evaluation metrics beyond accuracy, including unexpectedness, serendipity, novelty and diversity \cite{shani2011evaluating}.

Previous research introduces multiple alternative definitions of these measures, the goal of which is to provide novel, surprising and not previously seen recommendations. For example in \cite{adamopoulos2015unexpectedness}, the authors define unexpectedness as s distance of an item from the set of expectations and show the proposed approach achieves strong recommendation performance. However, one problem with this approach is that the set of expected items is defined in the limited sense as a closure of a set of previously consumed items, while a more comprehensive approach would look into the latent, complex and heterogeneous relations between users, items and entities and form the unexpectedness accordingly. These relations can be modeled using the concept of Heterogeneous Information Network (HIN) \cite{sun2013mining} that contains multiple types of objects and multiple types of links within a single network.

However to compute unexpectedness, it is hard to define the distance from the set of expected items in the HIN due to its discrete and complicated structure. In addition, latent relations between users and items are missing in the model, as it is not sufficient to accurately provide recommendations using only the explicit relations between users and items \cite{zhang2017deep}.

Therefore to address these problems, we propose to define the unexpectedness as the distance of an item from the expected item sets not in the \textit{feature space} (features and attributes of users and items) but in the \textit{latent space} (feature and attribute embeddings). We utilize the heterogeneous random walk mechanism to obtain the network embeddings of HIN. Then we define the unexpectedness as the euclidean distance from item embeddings to the \textit{latent convex hull} of the embeddings of the expected items. This approach has several advantages, including the guarantee of feasibility of the recommendation optimization and match with cognition theory \cite{gardenfors2004conceptual} of conceptual space, as we will describe in detail in Section 3. 

In this paper, we make the following contributions:

(1)We propose to apply deep-learning based method to the unexpected recommendation task and define unexpectedness in the latent space rather than in the feature space.

(2)We propose to formulate expectations of a user as a convex hull generated by all the previously consumed items in the latent space. This convex hull approach has strong theoretical foundations as shown in \cite{zenker2015applications}.

(3)Unlike the previously proposed approaches, we model the set of expectations in the feature space of HIN, which captures the complex and heterogeneous relations between users, items and entities. We subsequently map the HIN into the latent space and construct the convex hull from the latent structure for unexpectedness computation.

(4)We conduct extensive experiments on two real-world datasets and show that the proposed method consistently and significantly outperforms the other baseline models and the state-of-the-art unexpected recommendation algorithms in terms of various unexpectedness metrics, while achieving the same level of recommendation accuracy. Also, our method achieves higher maximum convex hull coverage than the baseline models and therefore recommends more semantically diverse items to the users.

The rest of the paper is organized as follows. We discuss the related work in Section 2 and present our proposed model in Section 3. Experimental design on the Yelp Dataset and TripAdvisor Dataset are described in Section 4 and the results as well as discussions are presented in Section 5. Finally, Section 6 summarizes our contributions and concludes the paper.

%% file: RelatedWork.tex
\section{Related Work}
In this section, we will introduce the prior literature on unexpectedness, alternative definition of expected set and state-of-the-art unexpected recommendation algorithms while pointing out the limitation of the previous models and comparing them with our proposed approach. We also describe the previous study on heterogeneous information network at the end of this section.

Researchers have addressed the importance of incorporating unexpectedness in recommendations \cite{kotkov2018investigating}, which could help overcome the overspecialization problem \cite{adamopoulos2015unexpectedness,iaquinta2010can}, broaden user preferences \cite{herlocker2004evaluating,zhang2012auralist,zheng2015unexpectedness} and increase user satisfaction \cite{adamopoulos2015unexpectedness,zhang2012auralist,lu2012serendipitous}. Note that, the concepts of unexpectedness and serendipity are closely related with each other, but still different in terms of definition and calculation. In particular, serendipity involves a positive emotional response of the user about a previously unknown item and measures how surprising these recommendations are \cite{shani2011evaluating}. 

Unexpectedness, on the other hand, measures the recommendations to users of those items that are not included in their consideration sets and depart from what they would expect from the recommender system. \cite{kontonasios2012knowledge} surveys different methods for discovering the unexpected patterns using frequent itemsets, tiles, association rules and classification rules; \cite{murakami2007metrics,ge2010beyond} defines unexpectedness as the deviation of a recommender system from the results obtained from a primitive prediction model; \cite{akiyama2010proposal} defines unexpectedness as an unlikely combination of item features; and \cite{adamopoulos2015unexpectedness} that defines unexpectedness as the distance of item from the set of expected items. However, these definitions do not consider the entity information in user reviews that bridge the expectation between users and items, which is crucial in modeling the expectation and preferences of certain users as pointed out in \cite{yu2013recommendation,shi2018heterogeneous,yu2014personalized}. Besides, these definitions determine the unexpectedness on the feature space, so they fail to capture the latent semantic relationship between users and items. In addition, these definitions only focus on the explicit correlation between users and items without considering the situation that the user could inference the expectation based on the historical behaviors. To address all the limitations, in this paper we propose to define the unexpectedness as the distance of item from the closure set of expected items for user in the latent space.


Research have also proposed various unexpected recommendation models, including Serendipitous Personalized Ranking\cite{lu2012serendipitous} that extends traditional personalized ranking methods by considering item popularity in AUC optimization; Auralist\cite{zhang2012auralist} that balances between the desired goals of accuracy, diversity, novelty and serendipity simultaneously; and HOM-LIN \cite{adamopoulos2015unexpectedness} that defines unexpectedness as the distance between items and the expected set of users. However as pointed out before, these models do not consider the latent interaction between users and items as well as the complexity and heterogeneous relations from heterogeneous entities, while the proposed Latent Convex Hull approach fits into the gap and achieves significantly better performance. We list the comparison between proposed model and the literature in Table \ref{compare}.

Another body of related work is around utilizing heterogeneous information network \cite{shi2017survey} and its embeddings for modeling complex heterogeneous context information and providing better recommendations. \cite{shi2018heterogeneous} transforms the learned node embeddings by a set of fusion functions and subsequently integrated into an extended matrix factorization model for the rating prediction task. \cite{han2018aspect} extracts different aspect-level similarity matrices of users and items through heterogeneous information network, and then feeds an deep neural network to learn aspect-level latent factors for recommendation. \cite{dong2017metapath2vec} formalizes meta-path-based random walks to construct the heterogeneous neighborhood of a node and then leverages a heterogeneous skip-gram model to perform node embeddings. However, all these previous work only focus on the usefulness and accuracy of recommendations, while failing to take unexpectedness into account, which is very important due to the previous literature \cite{kotkov2018investigating}.

\begin{table}[ht]
\centering
\begin{tabular}{c|ccccc} \hline
Algorithms                & LCH     & SPR      & Auralist & HOM-LIN & Random  \\ \hline
Latent Embeddings   & \cmark  & \xmark & \xmark & \xmark & \xmark \\
HIN & \cmark  & \xmark & \xmark & \xmark & \xmark \\
User Reviews           & \cmark  & \xmark & \cmark & \cmark & \xmark \\
Domain Knowledge   & \cmark   & \xmark & \cmark & \cmark & \xmark \\
Past Transactions     & \cmark   & \cmark & \cmark & \cmark & \xmark \\
Ratings                     & \cmark  & \cmark & \cmark & \cmark & \xmark \\ \hline
\end{tabular}
\newline
\caption{Comparison of Unexpected Recommendation Methods}
\label{compare}
\end{table}

%% file: Method.tex
\section{Model}
In this section, we describe our proposed Latent Convex Hull (LCH) model and unexpected recommendation algorithms. We will introduce the definition, sources, modeling and understanding of the unexpectedness, the setup of the feature space, intuition and advantages of using latent convex hull, the mapping from the feature space to the latent space and the unexpected utility function based on the proposed definition.

\subsection{Definition of Unexpectedness}
Following the prior literature \cite{adamopoulos2015unexpectedness}, the definition of unexpectedness starts with the modeling of the ''expected set'', the set of items that the user either previously encountered or closely related to them. Intuitively, users should have zero unexpectedness with respect to the items they have visited, purchased or rated before.  It is worth noticing, however, that the expected items could be more than that because of the various interactions and relatedness between users and items. In particular the set of expected items contains those that either viewed by the user or could be expected by the complex relations with those items that the user has viewed before. The closure of the ''expected'' items forms the ''expected set'' of the user, and we define unexpectedness as the distance of an item from the closure of the set of expected items for user.

\begin{figure*}
\centering
\begin{subfigure}{.5\textwidth}
  \centering
  \includegraphics[width=\textwidth]{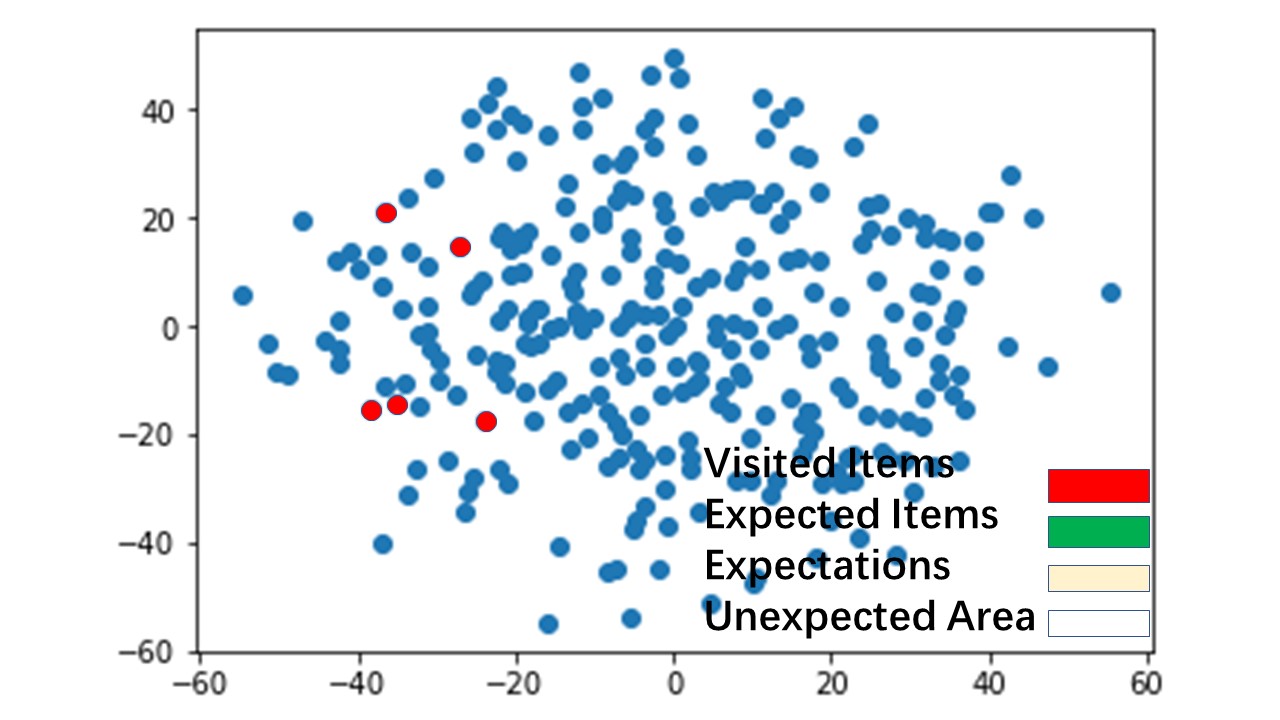}
  \caption{Visualization of Latent Space}
  \label{fig1:sub1}
\end{subfigure}%
\begin{subfigure}{.5\textwidth}
  \centering
  \includegraphics[width=\textwidth]{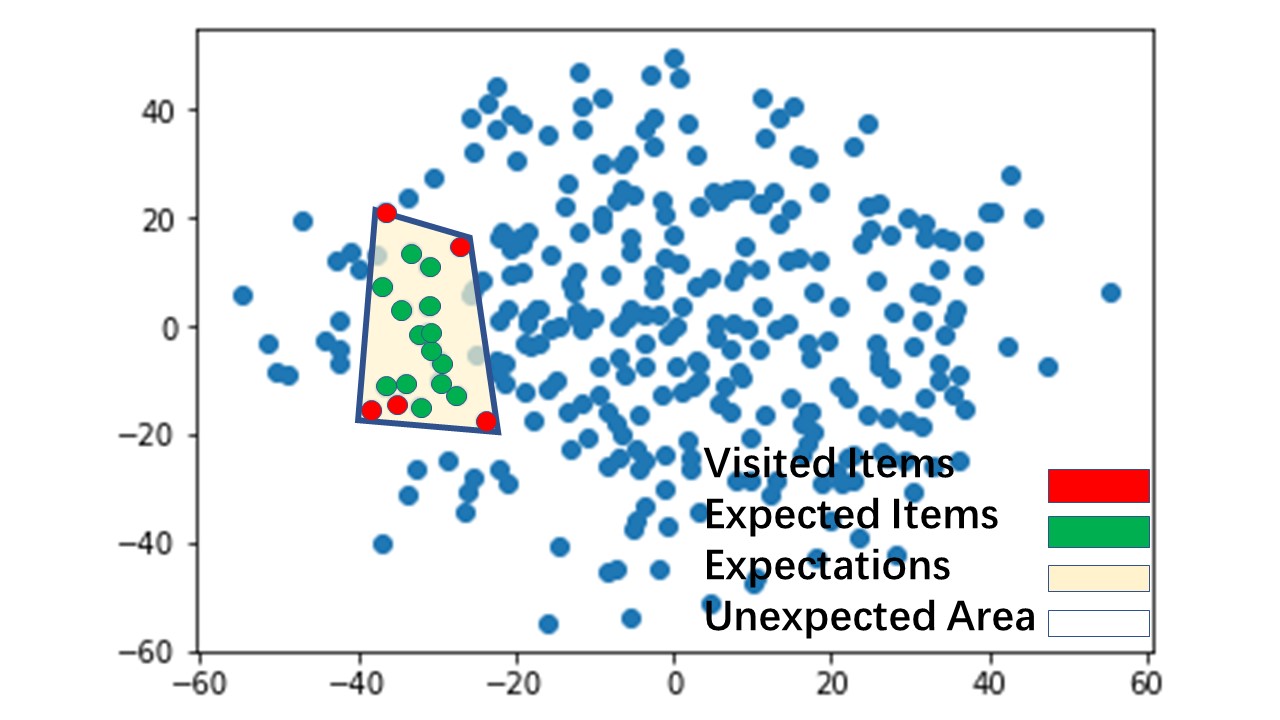}
  \caption{Visualization of Latent Convex Hull}
  \label{fig1:sub2}
\end{subfigure}
\caption{Visualization}
\label{definition}
\end{figure*}

Note that, we can define the unexpectedness in the feature space or in the latent space using the same approach. However, due to the discrete and complicated structure, it is difficult to model the expected items in the feature space. Therefore we propose to define the unexpectedness in the latent space, as we will describe in detail in the following section. We visualize the definition of these concepts in the latent space in Figure \ref{fig1:sub1} and \ref{fig1:sub2}. 
 
\subsection{Feature Space: Heterogeneous Information Network}
\begin{figure}
\centering
\includegraphics[width=0.5\textwidth]{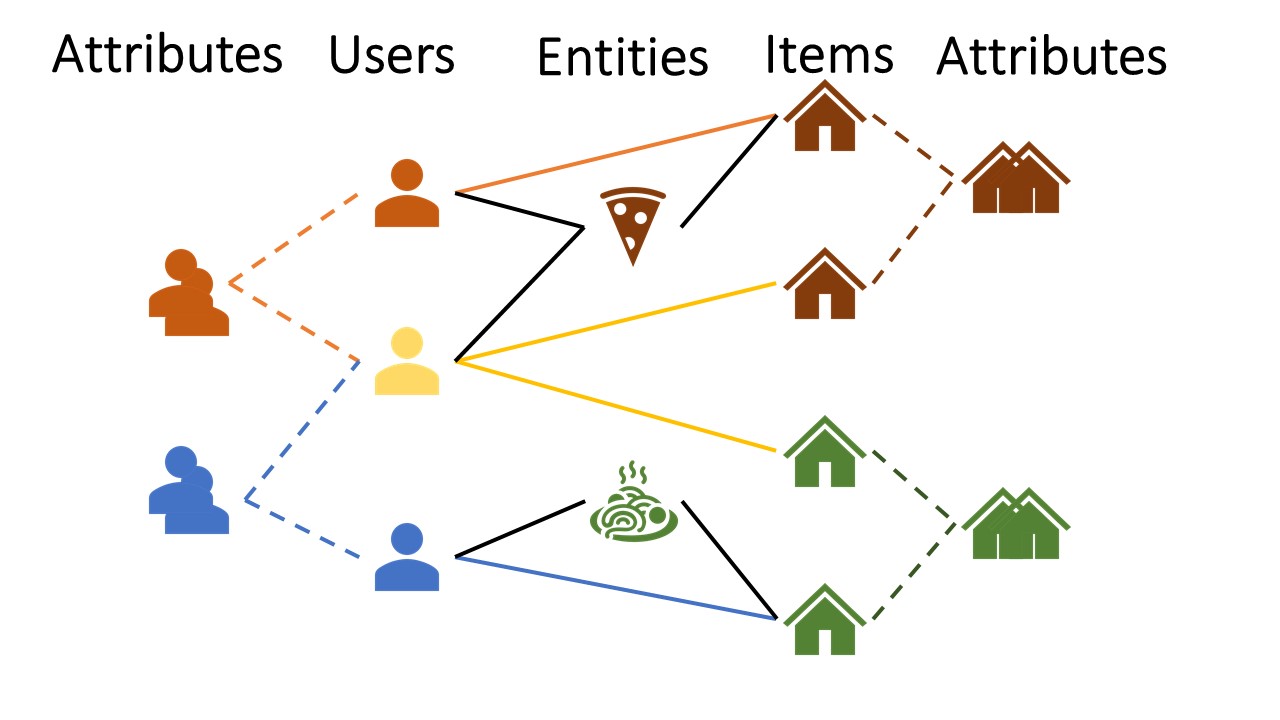}
\caption{Visualization of HIN}
\label{HIN}
\end{figure}

In order to model the set of expectations of the user, it is important to select an appropriate data structure capturing the expected relations between users, items and entities in the feature space. Intuitively, in the case of restaurant recommendations, the customer might get the expectation of certain restaurant because the customer (1) has visited that restaurant before, (2) has visited restaurants that are very similar (e.g., of the same franchise or of the same category), (3) has enjoyed the same cuisines served in that restaurant at other places and (4) gets to know that restaurant from the friends. This suggests that we should consider not only the direct interactions between users and items, but also the intermediate information from certain attributes and entities simultaneously. To capture the complex and multi-dimensional relations in the data record, we propose to use heterogeneous information network (HIN) \cite{sun2013mining} that contains multiple types of objects and multiple types of links in a single network. Specifically, the heterogeneous information network includes users, items, transactions, ratings, entities extracted from reviews and the meta-data information. We link the associated entities with corresponding users and items in the network. As an example, Figure \ref{HIN} demonstrates HIN for the restaurant application and show the relations between users, items and entities.

\subsection{Latent Space: Network Embeddings}
Note that, due to the discrete and complicated structure of heterogeneous information network, it's very hard to properly define the concept of ''unexpectedness'' and the distance metric on the feature space. Motivated by the goal to capture latent semantic interactions between users and items, we will introduce the deep-learning based network embedding approach in this section.

To learn effective node representations for the heterogeneous network $G=(V,E,T)$, following the setup in \cite{dong2017metapath2vec}, we enable the skip-gram mechanism to maximize the probability of having the heterogeneous context $N_{t}(v)$, $t\in T_{v}$ given a node $v$:
\[argmax_{\theta}\sum_{v \in V}\sum_{t \in T_{v}}\sum_{c_{t} \in N_{t}(v)}log P(c_{t}|v;\theta )\]
where $N_{t}(v)$ denotes the neighborhood of $v$ with the $t$th type of nodes and $P(c_{t}|v;\theta )$ defines the conditional probability of having a context node $c_{t}$ given a node $v$. To transform the structure of heterogeneous information network into skip-grams for optimization, we follow the natural idea of heterogeneous random walk to generate paths of multiple types of nodes in the network. Specifically, given a heterogeneous information network $G=(V,E,T)$, we generate the meta-path scheme as a path that is denoted in the form of $V_{1} \xrightarrow{R_{1}} V_{2} \xrightarrow{R_{2}} V_{3} \cdots V_{n} $ wherein $R = R_{1} \circ R_{2} \circ \cdots R_{n} $ defines the composite relations between the start and the end of the heterogeneous random walk. The transition probability within each random walk between two nodes is defined as follows:
\[p(V_{t+1}| V_{t}) = \begin{cases}
    \frac{C(T_{V_{t}},T_{V_{t+1}})}{|N_{t+1}(V_{t})|},& (V_{t}, V_{t+1}) \in E \\
    0,              & (V_{t}, V_{t+1}) \notin E
\end{cases}
\]
where $C(T_{V_{t}},T_{V_{t+1}})$ stands for the transition coefficient between the type of node $V_{t}$ and the type of node $V_{t+1}$. In our user-item-entity heterogeneous information network, we have 6 different transition coefficients which sums to one: $C_{UU}, C_{UE}, C_{UI}, C_{EI}, C_{EE}, C_{II}$ $|N_{t+1}(V_{t})|$ stands for the number of nodes of type $V_{t+1}$ in the neighborhood of $V_{t}$. In the heterogeneous information network, we perform heterogeneous random walk starting from each node iteratively and get the collection of meta-paths.

Note that, there are several benefits of utilizing heterogeneous random walks over other graph traversing approaches in HIN. First, heterogeneous random walks are computationally efficient in terms of both space and time requirements. In addition, heterogeneous random walk increases the effective sampling rate by reusing samples across different source nodes as it imposes graph connectivity in the sample generation process. And finally, it provides us a convenient way to address the heterogeneous influence of different types of nodes and links in HIN, as we can apply optimization algorithm to learn the transition coefficients efficiently. In this way, the embeddings of users and items could be obtained from the skip-gram mechanism \cite{mikolov2013distributed} on the meta-paths of heterogeneous random walk.

\subsection{Latent Convex Hull as Expected Set}
As explained before, we choose to take the closure in the latent space rather than original feature space. We utilize the well-defined concept \textit{convex hull} as a natural closure of the expected item embeddings. This approach provides the following advantages:

(1)The convexity property guarantees the feasibility of the recommendation as an optimization problem. Note that in our setting, the objective function, i.e., the utility function will be a linear combination of the rating and unexpectedness measures, so the convexity of the ''expected set'' will automatically imply the convexity of the objective function. The domain set of the optimization is a finite set of available items, so by Slater's Condition \cite{slater2014lagrange}, the primal problem is guaranteed feasible.  

(2)The convex hull corresponds to the cognition theory of conceptual space \cite{gardenfors2004conceptual}, a geometric structure that represents a number of quality dimensions that denote basic features by which concepts and objects can be compared, such as weight, color, taste, temperature, pitch, and the three ordinary spatial dimensions.. In the application of unexpected recommender system, the conceptual space includes ratings that users given to items and unexpectedness that measures the familiarity of users to items. According to the research in \cite{zenker2015applications}, natural categories are convex regions in conceptual spaces, therefore it is natural to model the closure as convex hull following the conceptual space theory.

(3)The convex hull could capture semantic interactions between users and items. Compared to the alternative definitions of an expected set, including Content-Based Similarity and Associate Rule Learning approach, the convex hull could utilize richer information to discover the relationship between users and items more precisely, including the intermediate effect from entities by the convex extension of expectation.

Based on these good properties, the proposed Latent Convex Hull model is a strong approach to define expected set for users and unexpectedness based on the expectations.


\subsection{Unexpected Recommendation: Latent Convex Hull}
Based on the network embeddings and the continuous structure of the latent space, we could now construct the expected set for each user using the proposed definition of unexpectedness. As described in the previous section, convex hull has certain advantages over other geometric structure to model the closure of expected set, so in this paper, we define the unexpectedness between each user/item pair as the distance between the item embedding and the latent convex hull generated from the user embedding and its neighbors. Specifically, we calculate the euclidean distance from the given point to the boundaries of the convex hull of the expected items. We assure that the euclidean distance is well-defined and unique by the hyperplane separation theorem \cite{}. Note that, the unexpectedness metric will take \textbf{negative} value if the given item is inside the convex hull (which means that the given item may lie deep within the user's expected set and the user could be overfamiliar with that item), and take positive value if it is indeed outside of the convex hull. The unexpectedness is formally defined below, where $\bar{user}$ stands for the latent convex hull generated by the user. 
\[Unexp(user, item) = d(item, \bar{user})\]

Once we set up the definition of unexpectedness, we could perform the unexpected recommendation based on the hybrid utility function: 
\[Utility(user, item)= (1-\alpha)*Rating(user, item) + \alpha *Unexp(user, item)\]

which incorporates the linear combination of ratings (which stand for usefulness) and unexpectedness. The key idea lies in that, instead of recommending the similar items that the users are very familiar with as the classical recommenders do, we wish to recommend unexpected and relevant items to the users that they might have not thought about, but indeed fit well to their satisfactions. Those two adversarial forces work together to get the optimal solution and thus get the best performance in terms of accuracy and unexpectedness measures. 

%% file: Experiments.tex
\section{Experiments}
To validate the superiority of our approach, we conduct extensive experiments on two distinctive real-world datasets and compare our methods with the state-of-the-art baseline recommendation models. In this section, we will introduce the two datasets, evaluation metrics and baseline models. Specifically, our experiments are designed to address the following research questions:

RQ1: How does the proposed definition of ''expected set'' perform compared to the alternative definitions?

RQ2: How does our model perform compared to the state-of-the-art unexpected recommendation models?

RQ3: Can our model reach a higher coverage of the latent space compared to other unexpected recommendation models?

\subsection{Datasets}
We conduct extensive experiments on two real-world datasets to evaluate the performance of our proposed model: the Yelp Challenge Dataset Round 12 \footnote{https://www.yelp.com/dataset/challenge}, which consists of 5,996,996 reviews from 1,518,169 users on 188,593 businesses on Yelp platform and also contains the category of restaurants, the friendship between users and information about time and location; the TripAdvisor Dataset \footnote{http://www.cs.cmu.edu/~jiweil/html/hotel-review.html}, which consists of 878,561 reviews from 576,689 users of 3,945 businesses on TripAdvisor platform. We list the descriptive statistics of these two datasets in Table \ref{statisticalnumber}. To address the cold-start issue, we filter out users and items that appear less than 5 times in the dataset.

\begin{table}
\centering
\begin{tabular}{|c|c|c|}
\hline
Dataset & \textbf{Yelp} & \textbf{TripAdvisor} \\ \hline
Number of Reviews & 5,996,996 & 878,561 \\ \hline
Number of Unique Businesses & 188,593 & 576,689 \\ \hline
Number of Unique Users & 1,518,169 & 3,945 \\ \hline
Average Reviews Per User & 4 & 2 \\ \hline
Average Reviews Per Business & 32 & 222 \\ \hline
\end{tabular}
\newline
\caption{Descriptive Statistics of Two Datasets}
\label{statisticalnumber}
\end{table}

\subsection{Evaluation Metrics}
To compare the performance of our proposed unexpected recommendation model and the baseline models, we follow \cite{herlocker2004evaluating} and measure its recommendation performance in terms of RMSE, MAE, Precision@N and Recall@N metrics. Besides, to measure the unexpected recommendation performance, we also compute Serendipity, Diversity and Coverage performance metrics following their definitions in \cite{ge2010beyond}: Serendipity = (RS\& PM)/PM, Diversity = (RS\& PM \&USEFUL)/PM, where RS stands for the recommended items using the selected model, PM stands for the recommendation results using a primitive prediction algorithm (usually selected as the linear regression) and USEFUL stands for the items whose utility is above certain threshold. Coverage \cite{ge2010beyond} is computed as the percentage of distinctive recommended items over all the distinctive items in the dataset.

\subsection{Baseline Models}
To validate the effectiveness of our proposed LCH modeling of the expected set, we compare it with the alternative approaches in terms of expected set and algorithms. The baseline models to compute the expected set include Base, CBS and ARL as described in \cite{adamopoulos2015unexpectedness}. In addition, we also compare to the RO (Rating Only) model that does not include the unexpectedness component. 
\begin{itemize}
\item \textbf{Base.} The set of expected recommendations consists only the set of items that she or he has already rated. In particular in the dataset, the item is also considered expected if the user has already rated certain item that belongs to the same franchise or brand.
\item \textbf{CBS.} (Content-Based Similarity) The set of expected recommendations consists the set of items in the Base set and the items that are sufficiently correlated with those measured by the semantic similarity between review texts.
\item \textbf{ARL.} (Associate Rule Learning) The set of expected recommendations consists the set of items in the Base set and the items that are closely related to those in the Base set. Specifically, two restaurants are related if they are in the same category or overlap more than half of their cuisines or overlap more than half of their customers.
\end{itemize}

Besides, we also implement several state-of-the-art unexpected recommendation models and compare their performance with LCH in terms of unexpectedness, serendipity, diversity and the convex hull coverage in the latent space. The baseline models include
\begin{itemize}
\item \textbf{SPR \cite{lu2012serendipitous}.} Serendipitous Personalized Ranking is a simple and effective method for serendipitous item recommendation that extends traditional personalized ranking methods by considering item popularity in AUC optimization, which makes the ranking sensitive to the popularity of negative examples.
\item \textbf{Auralist \cite{zhang2012auralist}.} Auralist is a personalized recommendation system that balances between the desired goals of accuracy, diversity, novelty and serendipity simultaneously. Specifically in the music recommendation, the authors combine Artist-based LDA recommendation with two novel components: Listener Diversity and Musical Bubbles. We adjust the algorithm to fit in our restaurant and hotel recommendation scenario.
\item \textbf{HOM-LIN \cite{adamopoulos2015unexpectedness}.} It is the state-of-the-art unexpected recommendation algorithm, where the author propose to define unexpectedness as the distance between items and the expected set of users. In our experiment, we select Hom-Lin as the baseline model, which obtains the best performance compared to other variations according to that paper.
\item \textbf{Random.} Random is the baseline model where we randomly recommend items to users without considering any information about the ratings, unexpectedness, utility and so on.
\end{itemize}

\section{Results}
In this section, we report the experiment results and give answers to the research questions in Section 4.

\begin{table*}
\centering
\small
\begin{tabular}{|c|c|c|c|c|c|c|c|c|c|c|} \hline
Dataset & Algorithm & Expected Set & RMSE & MAE & Precision@5 & Recall@5 & Unexpectedness & Serendipity & Diversity & Coverage \\ \hline
\multirow{25}{*}{Yelp} & \multirow{5}{*}{FM} & RO & 0.9197 & 0.6815 & 0.7699 & 0.6123 & -0.0326 & 0.0978 & 0.0135 & 0.5369 \\
                                      &      & CH & 0.9233 & 0.6860 & 0.7642 & 0.6137 & \textbf{0.0377*} & \textbf{0.2203*} & \textbf{0.1122*} & 0.5443 \\
                                      &      & Base & 0.9389 & 0.7178 & 0.7371 & 0.5880 & -0.0002 & 0.1223 & 0.0808  & 0.5430 \\
                                      &       & ARL & 0.9364 & 0.7607 & 0.7083 & 0.5833 & 0.0102 & 0.1030 & 0.0906  & 0.5482 \\
                                      &       & CBS & 0.9384 & 0.7552 & 0.7282 & 0.5830 & 0.0079 & 0.1232 & 0.0928  & 0.5482 \\ \cline{2-11}
                       & \multirow{5}{*}{CoCluster} & RO & 0.9509 & 0.7153 & 0.7239 & 0.5909 & -0.0369 & 0.1819 & 0.0508 & 0.5830 \\
                                      &             & CH & 0.9631 & 0.6968 & 0.7237 & 0.5967 & \textbf{0.0447*} & \textbf{0.2278*} & \textbf{0.1224*} & \textbf{0.7601*} \\
                                      &             & Base & 0.9795 & 0.7334 & 0.7211 & 0.5941 & 0.0003 & 0.1370 & 0.0908  & 0.5393 \\
                                      &             & ARL & 0.9764 & 0.7607 & 0.7083 & 0.5833 & 0.0104 & 0.1030 & 0.0925  & 0.5482 \\
                                      &             & CBS & 1.0425 & 0.8069 & 0.7359 & 0.5490 & 0.0104 & 0.1204 & 0.0925  & 0.5482 \\ \cline{2-11}
                       & \multirow{5}{*}{SVD} & RO & 0.9132 & 0.7069 & 0.7680 & 0.5983 & -0.0346 & 0.1294 & 0.0395 & 0.5424 \\
                                     &        & CH & 0.9263 & 0.7094 & 0.7639 & 0.6112 & \textbf{0.0351*} & \textbf{0.2326*} & \textbf{0.1126*} & 0.5351 \\
                                      &      & Base & 0.9479 & 0.7433 & 0.7640 & 0.5755 & 0.0020 & 0.0999 & 0.0908  & 0.5424  \\
                                      &             & ARL & 0.9359 & 0.7303 & 0.7605 & 0.5888 & 0.0079 & 0.0566 & 0.0987  & 0.5424 \\
                                      &             & CBS & 1.0152 & 0.7803 & 0.7632 & 0.5344 & 0.0056 & 0.0876 & 0.0728  & 0.5535 \\ \cline{2-11}
                       & \multirow{5}{*}{NMF} & RO & 0.9526 & 0.7171 & 0.7249 & 0.5852 & -0.0350 & 0.1909 & 0.0547 & 0.5959 \\
                                     &        & CH & 0.9632 & 0.6973 & 0.7165 & 0.5852 & \textbf{0.0447*} & \textbf{0.2343*} & \textbf{0.1206*} & 0.7638 \\ 
                                      &      & Base & 0.9889 & 0.7772 & 0.7171 & 0.5788 & 0.0037 & 0.1403 & 0.0699  & 0.5830 \\
                                      &             & ARL & 0.9793 & 0.7636 & 0.7105 & 0.5795 & 0.0102 & 0.1029 & 0.0728  & 0.5774 \\
                                      &             & CBS & 1.0388 & 0.8040 & 0.7322 & 0.5442 & 0.0125 & 0.1207 & 0.0896  & 0.5482 \\ \cline{2-11}
                       & \multirow{5}{*}{KNN} & RO & 0.9123 & 0.7048 & 0.7688 & 0.6085 & -0.0336 & 0.0977 & 0.0130 & 0.5369 \\
                                     &        & CH & 0.9251 & 0.7060 & 0.7632 & 0.6136 & \textbf{0.0367*} & \textbf{0.2103*} & \textbf{0.1022*} & 0.5443 \\
                                      &      & Base & 0.9476 & 0.7443 & 0.7685 & 0.5805 & -0.0004 & 0.1043 & 0.0834  & 0.5442 \\
                                      &      & ARL & 0.9352 & 0.7300 & 0.7702 & 0.5985 & 0.0079 & 0.0611 & 0.0856  & 0.5480 \\
                                      &       & CBS & 1.0143 & 0.7794 & 0.7710 & 0.5451 & 0.0102 & 0.0885 & 0.0724  & 0.5461 \\ \hline
\multirow{25}{*}{TripAdvisor} & \multirow{5}{*}{FM} & RO & 1.1105 & 0.8340 & 0.6768 & 0.9590 & -0.0922 & 0.3979 & 0.0017 & 0.1798 \\
                              &                     & CH & 1.1275 & 0.8445 & 0.7040 & 0.9656 & \textbf{0.0643*} & \textbf{0.4631*} & \textbf{0.0493*} & 0.1798 \\
                              &                      & Base & 1.1550 & 0.8452 & 0.6772 & 0.8715 & -0.0266 & 0.4591 & 0.0301 & 0.1807 \\ 
                              &                      & ARL & 1.1512 & 0.8323 & 0.6803 & 0.9001 & 0.0097 & 0.4501 & 0.0365  & 0.1802 \\
                              &                      & CBS & 1.1343 & 0.8392 & 0.6809 & 0.9065 & 0.0122 & 0.4485 & 0.0332  & 0.1802 \\ \cline{2-11}
                              & \multirow{5}{*}{CoCluster} & RO & 1.0178 & 0.7643 & 0.6845 & 0.9732 & -0.0934 & 0.3973 & 0.0015 & 0.1855 \\
                              &                      & CH & 1.0511 & 0.8048 & 0.6947 & 0.9692 & \textbf{0.0652*} & \textbf{0.4619*} & \textbf{0.0471*} & 0.1798 \\
                              &                      & Base & 1.0657 & 0.8452 & 0.6917 & 0.8715 & -0.0266 & 0.4393 & 0.0210 & 0.1807 \\ 
                              &                      & ARL & 1.0577 & 0.8220 & 0.6801 & 0.9103 & 0.0179 & 0.4401 & 0.0371  & 0.1802 \\
                              &                      & CBS & 1.0573 & 0.8292 & 0.6902 & 0.9077 & 0.0122 & 0.4423 & 0.0302  & 0.1802 \\ \cline{2-11}
                              & \multirow{5}{*}{SVD} & RO & 0.9868 & 0.7533 & 0.7210 & 0.9465 & -0.0931 & 0.3967 & 0.0006 & 0.1798 \\
                              &                      & CH & 1.0214 & 0.7890 & 0.7182 & 0.8911 & \textbf{0.0644*} & \textbf{0.4621*} & \textbf{0.0499*} & 0.1798 \\
                              &                      & Base & 1.0368 & 0.8216 & 0.7087 & 0.8099 & -0.0262 & 0.4594 & 0.0298 & 0.1807 \\ 
                              &                      & ARL & 1.0354 & 0.8079 & 0.6992 & 0.8227 & 0.0009 & 0.4499 & 0.0333 & 0.1802 \\
                              &                      & CBS & 1.0345 & 0.7998 & 0.6999 & 0.8385 & 0.0207 & 0.4487 & 0.0366 & 0.1802 \\ \cline{2-11}
                              & \multirow{5}{*}{NMF} & RO & 1.0241 & 0.7709 & 0.6850 & 0.9681 & -0.0927 & 0.3979 & 0.0010 & 0.1798 \\
                              &                      & CH & 1.0575 & 0.8111 & 0.6869 & 0.9655 & \textbf{0.0644*} & \textbf{0.4627*} & \textbf{0.0499*} & 0.1798 \\
                              &                      & Base & 1.0672 & 0.8463 & 0.6922 & 0.8723 & -0.0270 & 0.4598  & 0.0261 & 0.1807 \\ 
                              &                      & ARL & 1.0552 & 0.8323 & 0.6902 & 0.9021 & 0.0109 & 0.4501 & 0.0365  & 0.5480 \\
                              &                      & CBS & 1.0543 & 0.8392 & 0.6969 & 0.9015 & 0.0222 & 0.4485 & 0.0334  & 0.5461 \\ \cline{2-11}
                              & \multirow{5}{*}{KNN} & RO & 0.9940 & 0.7531 & 0.6969 & 0.9689 & -0.0933 & 0.3979 & 0.0019 & 0.1798 \\
                              &                      & CH & 1.0275 & 0.7945 & 0.7040 & 0.9256 & \textbf{0.0643*} & \textbf{0.4631*} & \textbf{0.0492*} & 0.1798 \\
                              &                      & Base & 1.0434 & 0.8279 & 0.7012 & 0.8318 & -0.0266 & 0.4593 & 0.0200 & 0.1802 \\ 
                              &                      & ARL & 1.0352 & 0.8000 & 0.7002 & 0.8985 & 0.0019 & 0.4511 & 0.0256  & 0.1802 \\
                              &                      & CBS & 1.0343 & 0.8094 & 0.7010 & 0.8451 & 0.0002 & 0.4585 & 0.0224  & 0.1802 \\ \hline
\end{tabular}
\newline
\caption{Validation of Unexpected Recommendation on the two datasets. ''RO'': Rating Only, 'LCH'': Latent Convex Hull, ''CBS'': Content-Based Similarity, ''ARL'': Associate Rule Learning, ''*'' stands for 95\% statistical significance}
\label{result}
\end{table*}

\subsection{RQ1: Comparison of Expected Sets}
To validate our proposed definition of unexpectedness, we compare the recommendation performance using alternative definition of expected sets introduced in \cite{adamopoulos2015unexpectedness} and corresponding unexpected distance. We also include the results of standard recommendation, i.e., using rating only (RO) for recommendation. In addition, to verify the robustness of the experimental settings, we conduct the cross-validation experiment using five popular collaborative filtering algorithms including k-Nearest Neighborhood approach (KNN) \cite{altman1992introduction}, the Singular Value Decomposition approach (SVD) \cite{sarwar2002incremental}, the Co-Clustering approach \cite{george2005scalable}, the Non-Negative Matrix Factorization approach (NMF) \cite{lee2001algorithms} and the Factorization Machine approach (FM) \cite{rendle2010factorization}. We conduct these experiments on two real-world datasets, resulting in 400 experiments in total. 

\begin{figure*}[!]
    \centering
    \begin{subfigure}[t]{0.33\textwidth}
        \centering
        \includegraphics[width=\textwidth]{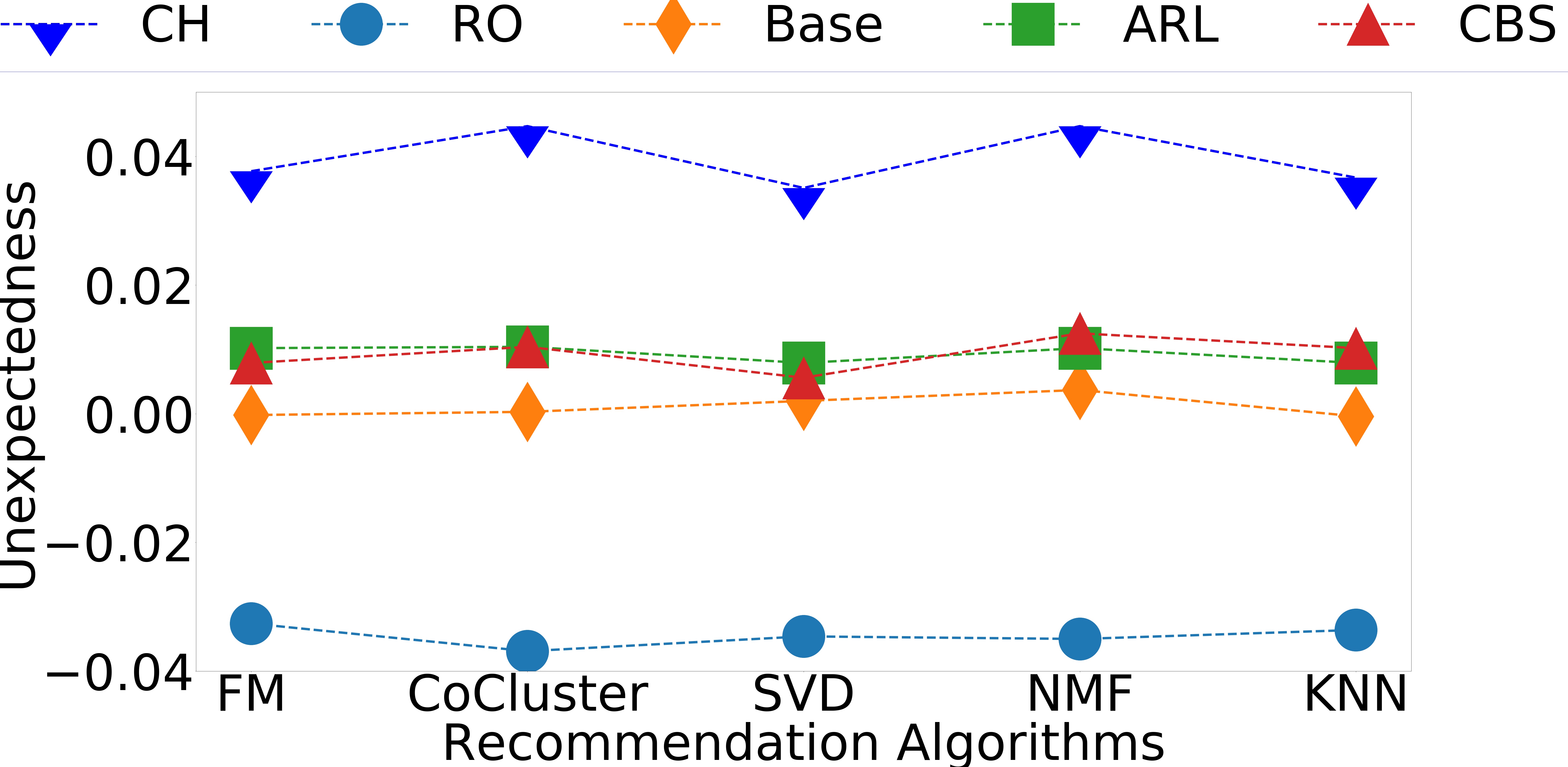}
        \caption{Unexpectedness}
    \end{subfigure}%
    ~ 
    \begin{subfigure}[t]{0.33\textwidth}
        \centering
        \includegraphics[width=\textwidth]{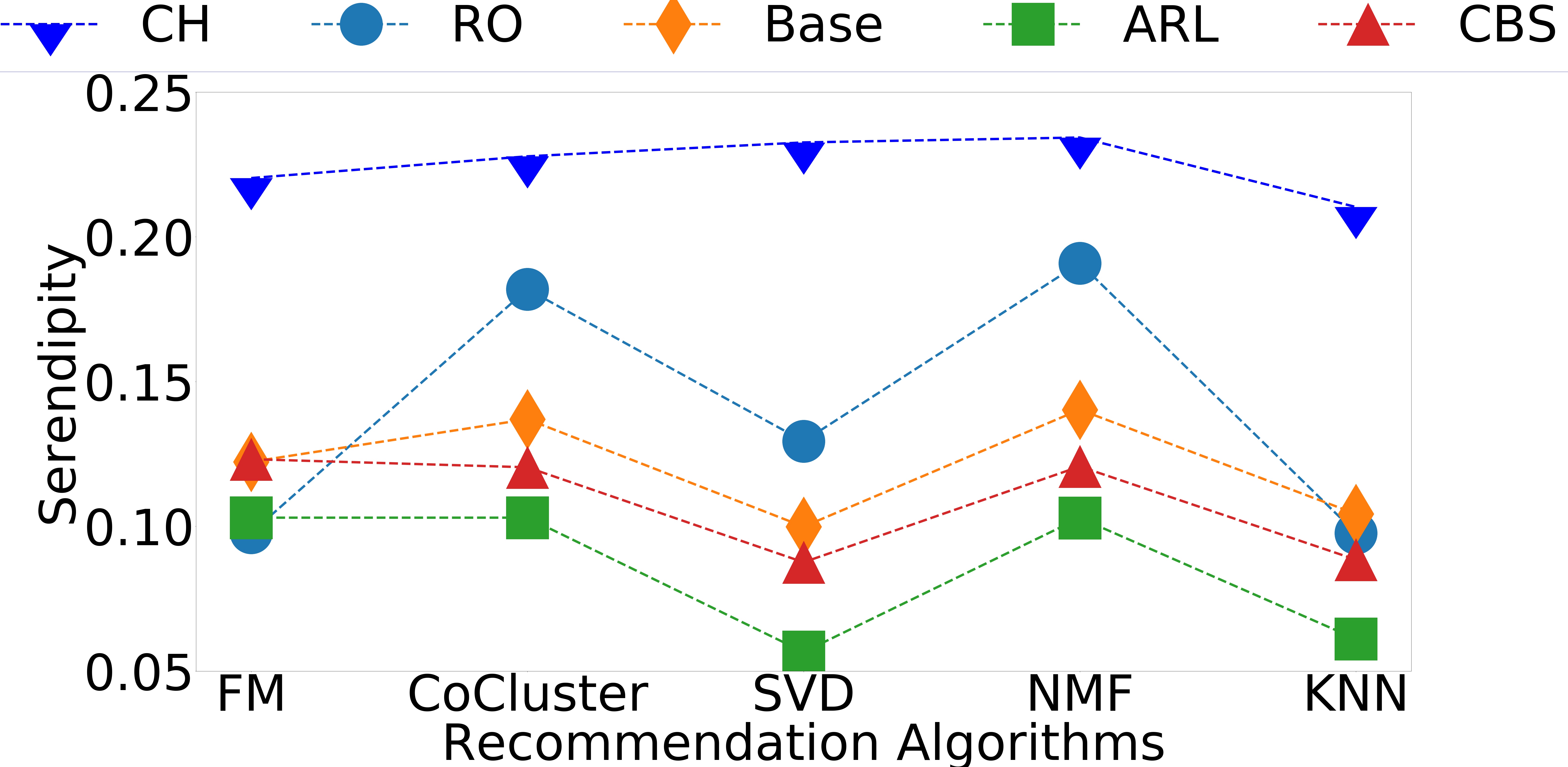}
        \caption{Serendipity}
    \end{subfigure}%
    ~ 
    \begin{subfigure}[t]{0.33\textwidth}
        \centering
        \includegraphics[width=\textwidth]{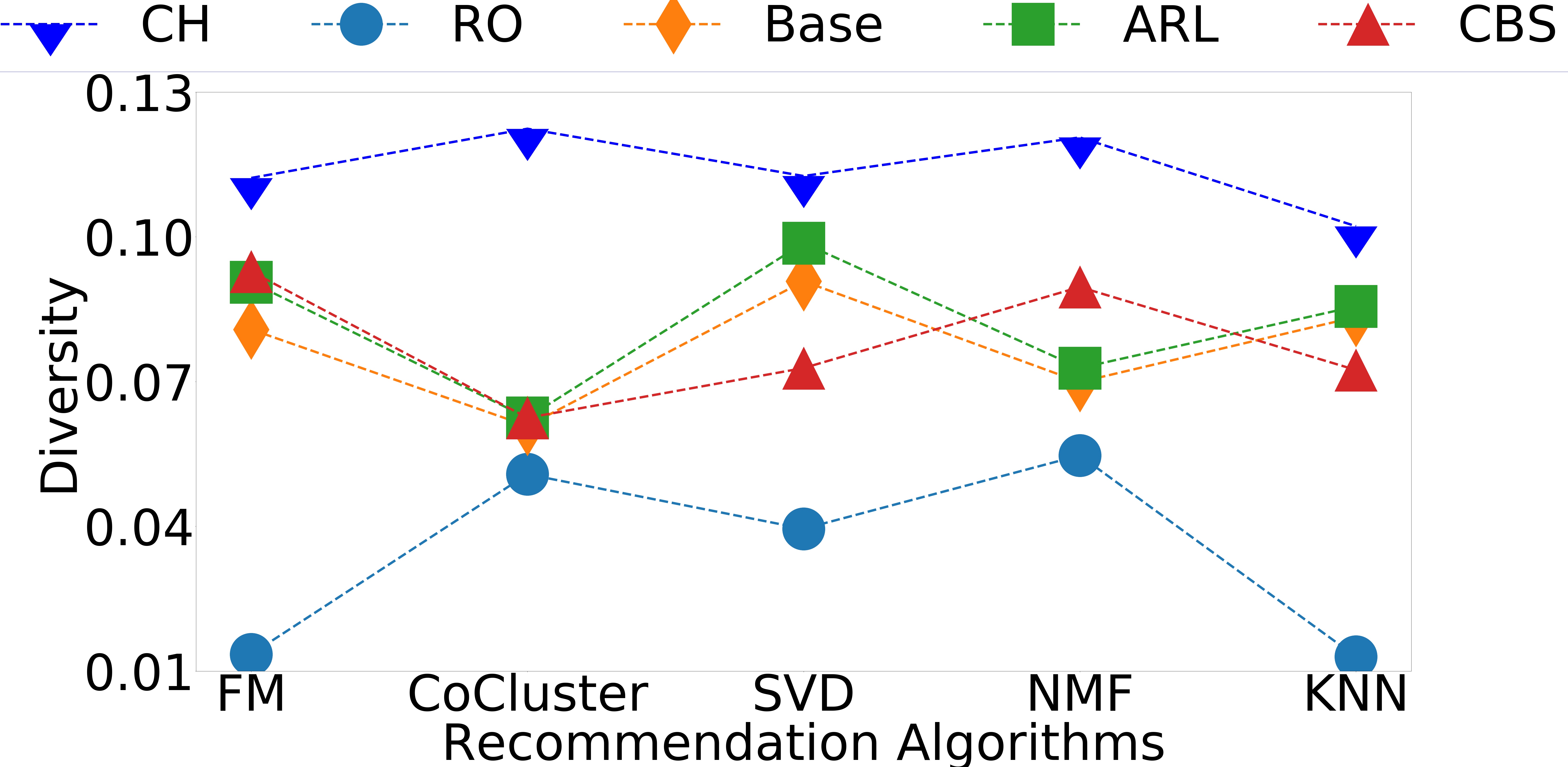}
        \caption{Diversity}
    \end{subfigure}
    \caption{Recommendation Performance of Yelp}
\label{Yelp}
\end{figure*}

\begin{figure*}[!]
    \centering
    \begin{subfigure}[t]{0.33\textwidth}
        \centering
        \includegraphics[width=\textwidth]{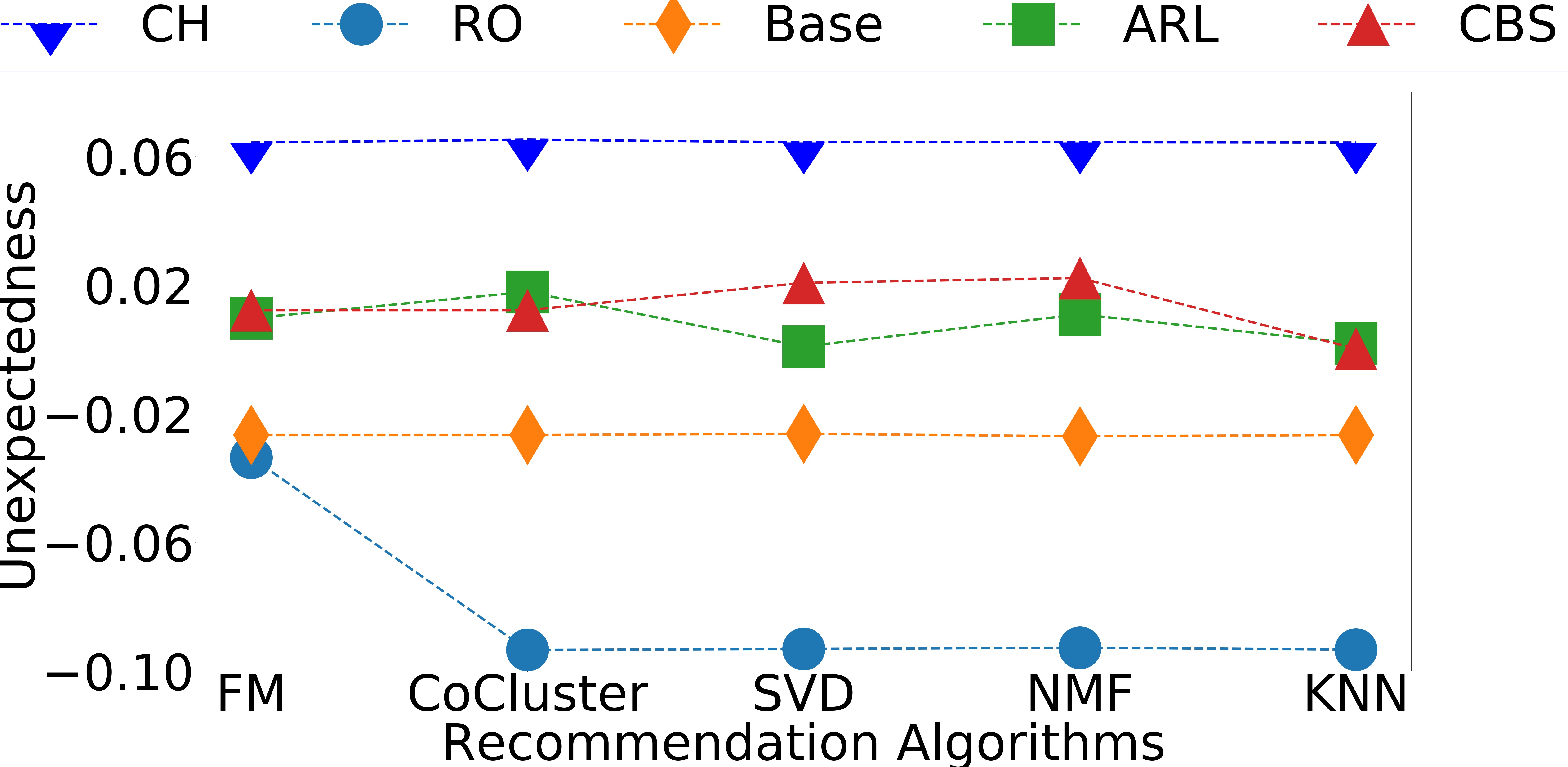}
        \caption{Unexpectedness}
    \end{subfigure}%
    ~ 
    \begin{subfigure}[t]{0.33\textwidth}
        \centering
        \includegraphics[width=\textwidth]{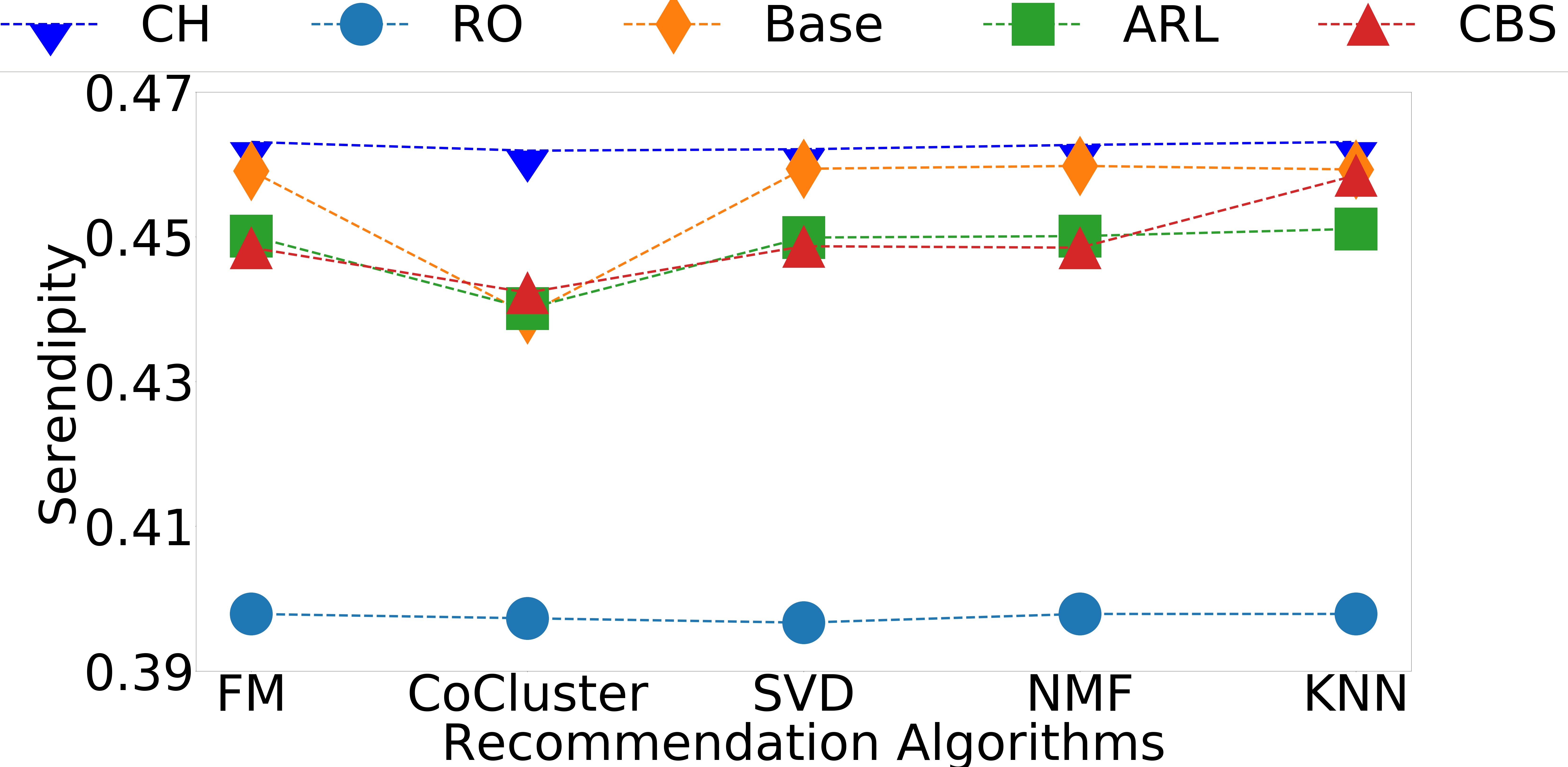}
        \caption{Serendipity}
    \end{subfigure}%
    ~ 
    \begin{subfigure}[t]{0.33\textwidth}
        \centering
        \includegraphics[width=\textwidth]{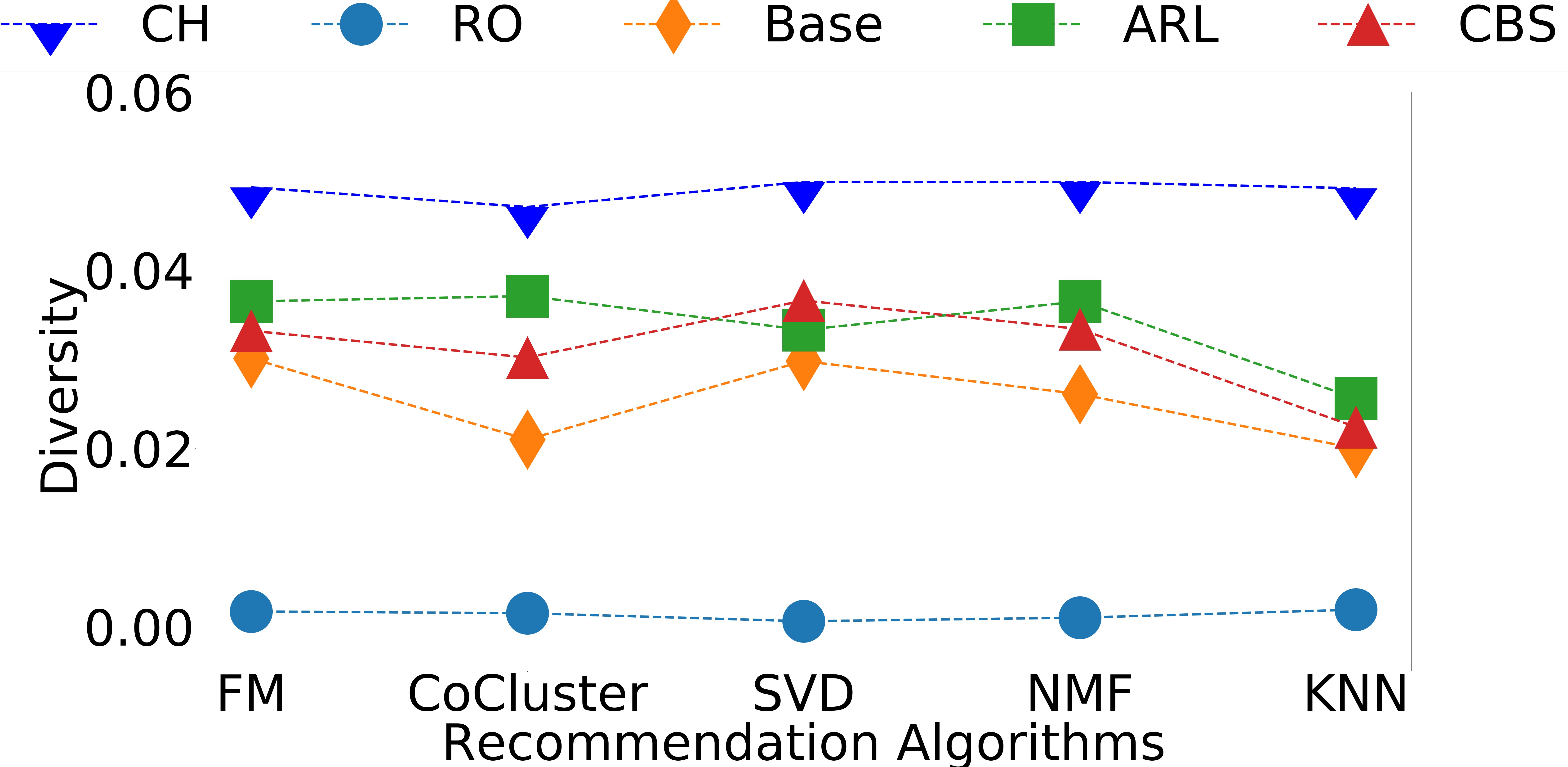}
        \caption{Diversity}
    \end{subfigure}
    \caption{Recommendation Performance of TripAdvisor}
\label{TripAdvisor}
\end{figure*}

The performance results are reported in Table \ref{result} and also in Figure \ref{Yelp} and \ref{TripAdvisor} that are based on Table \ref{result}. The results show that our proposed model consistently and significantly outperforms the baseline models over all the experimental settings. In paricular our model significantly increases the serendipity, unexpectedness and diversity measures, while still performing as good as the baseline models in terms of accuracy measures including RMSE, MAE, Precision and Recall. More specifically, we observe over 100\% increase in unexpectedness, 80\% increase in serendipity and 20\% increase in diversity measures on average, while the differences between the proposed and baseline models are statistically insignificant in terms of RMSE, MAE. Precision and Recall measures. To sum up, the answer to RQ1 is that our proposed definition of ''expected set'' using Latent Convex Hull approach performs consistently and significantly better than all other baseline methods.

It is also worth noting that some of the baseline models obtain negative values of unexpectedness, as reported in Table \ref{result}. Based on our definition of unexpectedness in the previous section, the metric will take \textbf{negative} value if the given item is inside the convex hull (which means that the user could be overfamiliar with that item), and take positive value if it lies outside of the convex hull. These negative values indicate that the alternative definitions of expected set suffer from the problem of filter bubble. Our proposed approach, however, achieves superior performance in terms of unexpectedness for all the experimental settings, which supports the claim that it is indeed a powerful tool to address the filter bubbles problem.

\subsection{RQ2: One-Time Recommendation}
To show that our model could indeed provide more unexpected recommendations than the state-of-the-art methods, we provide a set of one-time recommendation for all the users in the dataset to compare the unexpectedness performance. In particular each user is recommended a set of 10 items based on the past transactions and we use the same measures in the previous section to evaluate the unexpected recommendation performance. The experiment results are reported in Table \ref{yelp_one} and \ref{tripadvisor_one}, which show that our proposed model consistently and significantly outperforms all other baselines in terms of Coverage, Unexpectedness, Serendipity and Diversity meaasures. To sum up, the answer to RQ2 would be that our proposed LCH model performs consistently and significantly better than all the other state-of-the-art unexpected recommendation models.

\begin{table}
\centering
\begin{tabular}{ccccc} \hline
Algorithms & Coverage & Unexpectedness & Serendipity & Diversity\\ \hline
LCH          &   \textbf{0.3524}          &  \textbf{0.4998}     &  \textbf{1.0}   &   \textbf{0.6208} \\
SPR            &  0.1697               & 0.4668         & 0.972 &  0.4532       \\
Auralist       &  0.1457               & 0.4663         & 0.9637&  0.6047       \\
HOM-LIN         &  0.1365               & 0.4251         & 0.8629 & 0.6000        \\
Random      &  0.1457              & 0.3733         & 0.8848 & 0.5763        \\ \hline
\end{tabular}
\newline
\caption{Comparison of Unexpected Performance:Yelp}
\label{yelp_one}
\end{table}

\begin{table}
\centering
\begin{tabular}{ccccc} \hline
Algorithms & Coverage & Unexpectedness & Serendipity & Diversity \\ \hline
LCH           &   \textbf{0.2597}     &  \textbf{0.5582}     &  \textbf{0.9969}   &   \textbf{0.864}  \\
SPR           &  0.1834               & 0.4739         & 0.9593 &  0.8175        \\
Auralist      &  0.1834               & 0.4728         & 0.9562 &  0.8553        \\
HOM-LIN         &  0.2144               & 0.4722         & 0.9629 & 0.8117        \\
Random     &  0.2173              & 0.3733         & 0.9468 & 0.835        \\ \hline
\end{tabular}
\newline
\caption{Comparison of Unexpected Performance:TripAdvisor}
\label{tripadvisor_one}
\end{table}

\subsection{RQ3: Maximum Convex Hull}
To answer RQ3 and compare the long-term unexpected performance of the recommendation model, we first need to define the concept of the maximum convex hull. The \textit{Maximum Convex Hull} for each user is defined as the convex hull of all the items in the dataset whose utlity is above certain threshold. Intuitively, the maximum convex hull constitues the upper bound of the expected set for the user. The coverage percentage for convex hull is computed as the ratio of the area of expectations over that of the maximum convex hull.

\begin{figure*}
\centering
\begin{subfigure}{.5\textwidth}
  \centering
  \includegraphics[width=\textwidth]{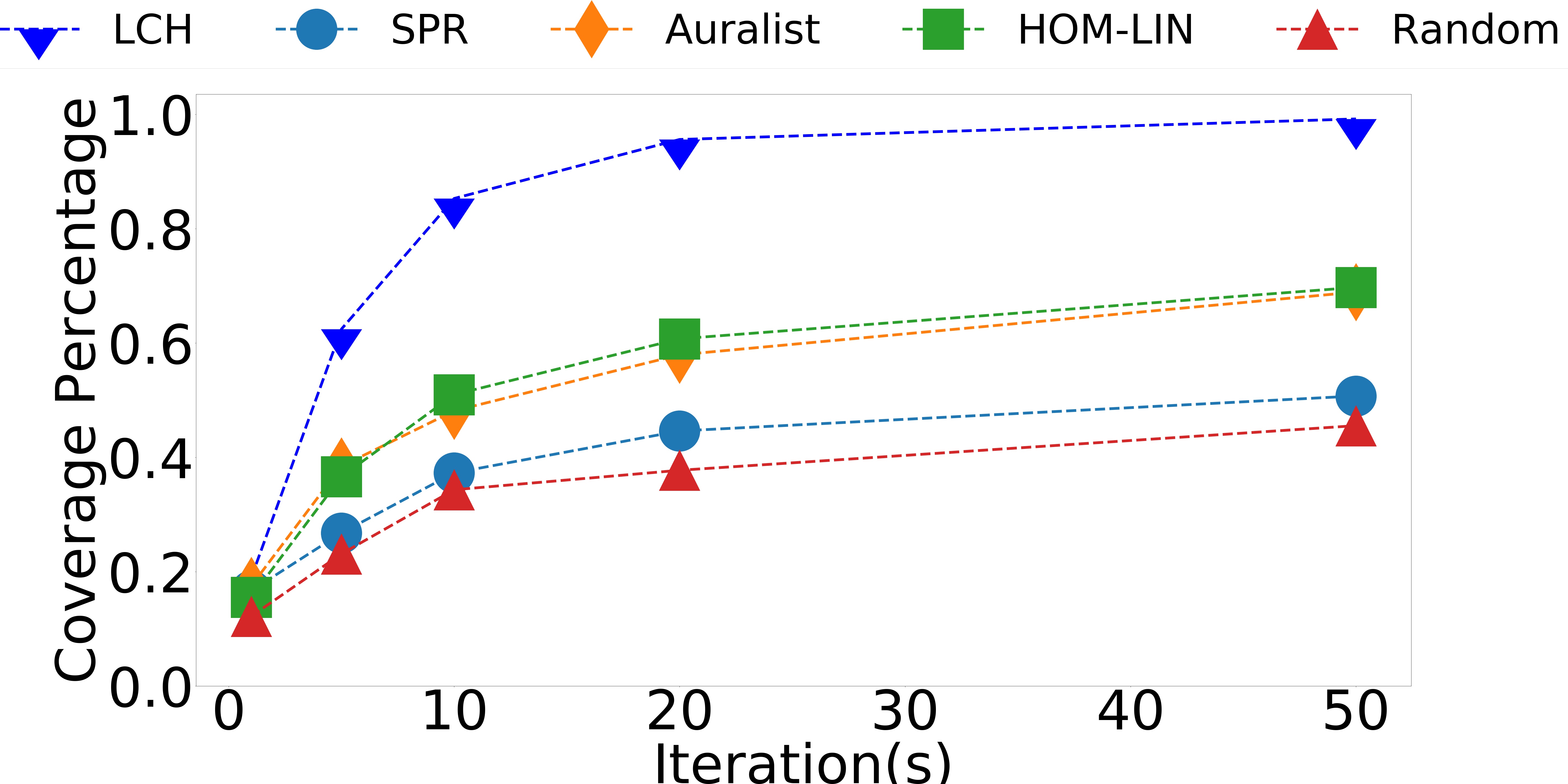}
  \caption{Yelp}
  \label{fig:yelp}
\end{subfigure}%
\begin{subfigure}{.5\textwidth}
  \centering
  \includegraphics[width=\textwidth]{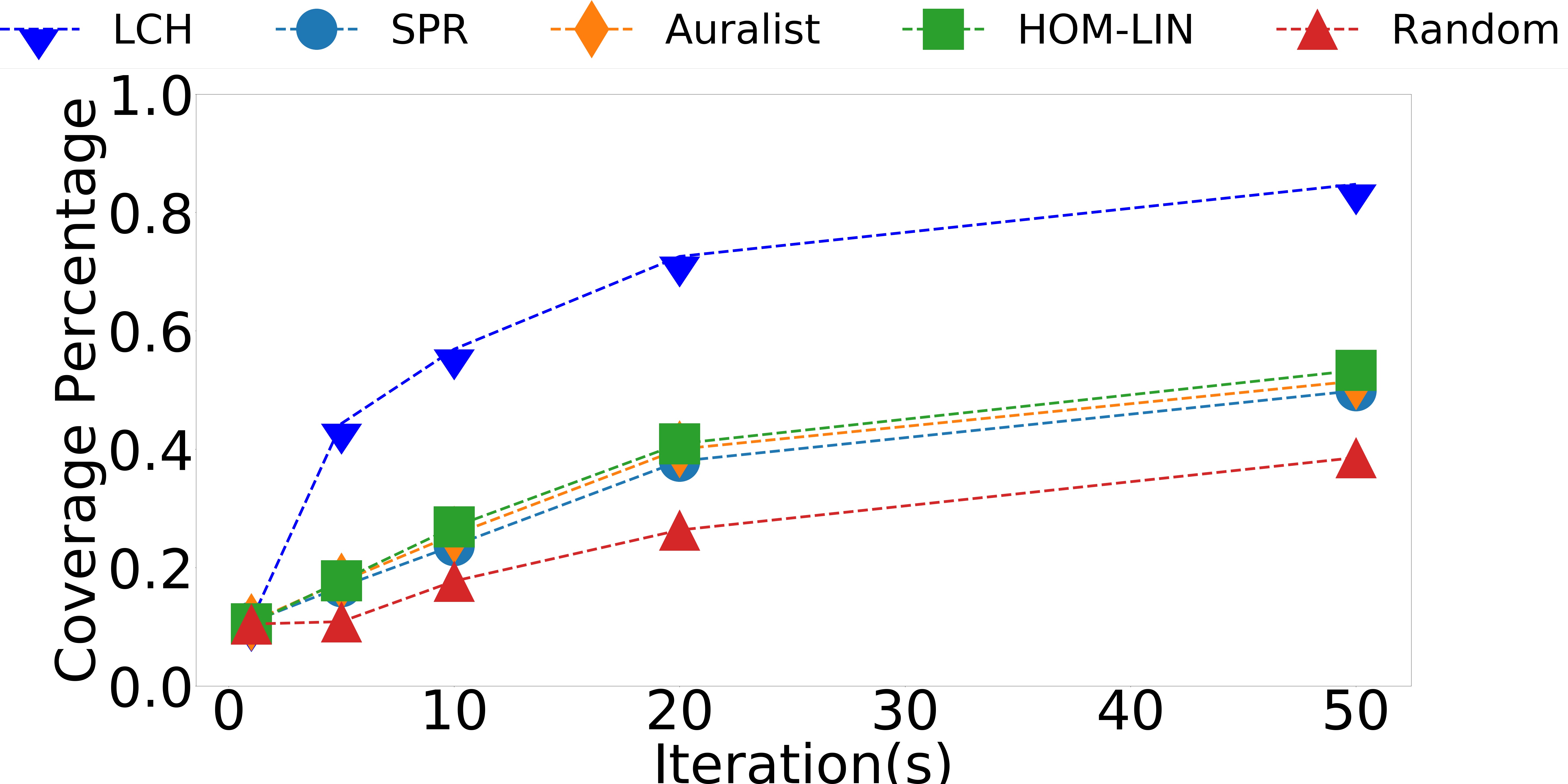}
  \caption{TripAdvisor}
  \label{fig:tripadvisor}
\end{subfigure}
\caption{Comparison of Latent Convex Hull Coverage}
\label{multiple}
\end{figure*}

As a part of our experiment, we repeat the item recommendation for users multiple times while observing how many novel recommendations have been generated during the iterative process. We are also interested to see if our proposed unexpected recommendation approach could reach the upper bound of the maximum convex hull for each user and how quick it extends the area of expected set. Furthermore when items are recommended in the previous iteration, it will become expected in the next iteration. We compare the coverage of the Maximum Convex Hull after 1,5,10,20,50 iterations, and the performance results are shown in Table \ref{yelp_multiple}, \ref{tripadvisor_multiple} and Figure \ref{multiple}. Figure \ref{multiple} shows that our LCH methods will significantly outperform the baseline models in terms of convex hull coverage over all the iteration situations. Moreover, the convergence rate to the maximum convex hull for the Yelp dataset is much faster than that of the TripAdvisor dataset. It happens because we have richer information for the restaurant recommendation: apart from user-item transaction, ratings and reviews, we also have the information about restaurant categories, cuisines and user friendship network, while we only have reviews and ratings data for the TripAdvisor dataset.

To sum up, we give the answers to all three research questions in Section 4 and conclude that the proposed LCH model achieves the best performance in unexpected recommendations compared to all the other state-of-the-art baseline models.

\begin{table}
\centering
\begin{tabular}{cccccc} \hline
Iterations & 1 & 5 & 10 & 20 & 50 \\ \hline
LCH &  \textbf{0.1894} & \textbf{0.6241} &  \textbf{0.8526} &  \textbf{0.9558}  &  \textbf{0.9917} \\
SPR            &        0.1667  & 0.2677         &       0.3728    &      0.4460     &      0.5072      \\
Auralist       &       0.1751  &  0.3843         &      0.4812    &      0.5792     &      0.6890    \\
HOM-LIN          &        0.1554 & 0.3660        &         0.5096     &      0.6072   &      0.6972  \\
Random       &      0.1219  & 0.2311         &        0.3432     &        0.3773   &        0.4553    \\ \hline
\end{tabular}
\newline
\caption{Comparison of Maximum Convex Hull Coverage:Yelp}
\label{yelp_multiple}
\end{table}

\begin{table}
\centering
\begin{tabular}{cccccc} \hline
Iterations & 1 & 5 & 10 & 20 & 50 \\ \hline
LCH &  \textbf{0.1101} & \textbf{0.4434} &  \textbf{0.5691} &  \textbf{0.7257}  &  \textbf{0.8477} \\
SPR            &        0.1081  & 0.1678         &       0.2374    &      0.3798     &      0.4989      \\
Auralist       &       0.1082  &  0.1765         &      0.2558    &      0.4002     &      0.5152    \\
HOM-LIN          &        0.1054 & 0.1781        &         0.2691     &      0.4094   &      0.5333  \\
Random       &      0.1047  & 0.1087         &        0.1772     &        0.2635   &        0.3859    \\ \hline
\end{tabular}
\newline
\caption{Comparison of Maximum Convex Hull Coverage:TripAdvisor}
\label{tripadvisor_multiple}
\end{table}

%% file: Conclusions.tex
\section{Conclusion}
In this paper, we propose a novel approach to provide unexpected and useful recommendations based on the concept of latent convex hull, which constitutes the convex closure set of expected items for the users. We define unexpectedness as the distance of an item in the latent space from the closure set of expected items for the user. We define the utility as a linear combination of unexpectedness and ratings and recommend items to the users based on this utility measure. Furthermore, we demonstrate that the proposed approach consistently and significantly outperforms other baseline models in terms of the unexpectedness, serendipity, diversity and coverage measures, which supports the validity and superiority of the LCH model.

The contributions of this paper are threefold. First, we propose a novel definition of unexpectedness based on the latent convex hull to capture the latent relationships between users and items via the embedding techniques and heterogeneous information network. Note that we define unexpectedness in the latent space, as opposed to the feature space, as have been done in all the previous proposed definitions of unexpectedness. Second, we propose an unexpected recommendation model based on this novel definition of unexpectedness. Specifically, the hybrid utility function is a linear combination of unexpectedness and usefulness. Third, we conduct extensive experiments and show that our proposed model consistently and significantly outperforms the other baseline models in terms of serendipity, unexpectedness, and diversity performance metrics, while achieving the same level of accuracy in terms of RMSE, MAE, Precision and Recall measures. We also show that our proposed model approaches the Maximum Convex Hull significantly faster than other models.

As the future work, we plan to conduct live experiments with real business environment in order to further evaluate the effectiveness of unexpected recommendations and analyze both qualitative and quantitative aspects in a traditional online retail setting, especially with the utilization of the A/B test. Moreover, we will further explore the convexity property of the user's expectations, which is introduced in Section 3. Specifically, we plan to connect cognitive psychology and field experiments to dig deeper into the theory. Finally, we plan to explore further the concept of unexpectedness \& relevance and investigate how to automatically combine the concept of unexpectedness into the deep-learning based recommender systems.